\documentclass{jors}

\usepackage[utf8]{inputenc}
\usepackage[T1]{fontenc}
\usepackage{dirtytalk}

\usepackage[super]{nth}
\usepackage{enumitem}
\usepackage{dirtytalk}
\usepackage{amsthm}
\usepackage{amsmath}
\usepackage{amssymb}
\usepackage{kpfonts}
\usepackage{dsfont}
\usepackage{hyperref}
\usepackage{graphicx}
\graphicspath{{Images/}}

\theoremstyle{definition}
\newtheorem{defn}{Definition}

\begin{document}
\rule{\textwidth}{1pt}

\section*{(1) Overview}

\vspace{0.5cm}

\section*{Title}

\textbf{The Hetero-functional Graph Theory Toolbox}

\section*{Paper Authors}

1. Thompson, Dakota J.$^{1,2,*}$ \\
2. Hegde, Prabhat$^1$ \\
3. Schoonenberg, Wester C. H.$^1$ \\
4. Khayal, Inas S.$^2$ \\
5. Farid, Amro M.$^1$

\section*{Paper Author Roles and Affiliations}
1. Thayer School of Engineering at Dartmouth, Hanover, NH, 03755, USA \\
2. The Dartmouth Institute Geisel School of Medicine at Dartmouth, Lebanon, NH, 03766, USA \\
$^*$Dakota.J.Thompson.TH@Dartmouth.edu

\section*{Abstract}

In the \nth{20} century, the analysis, design, planning, and operation of large-scale heterogeneous engineering systems from a holistic perspective has necessitated ever-more sophisticated modeling techniques.  Hetero-Functional Graph Theory (HFGT) has emerged as a means to address the complexity of engineering systems.  This recently developed \emph{Hetero-functional Graph Theory Toolbox} facilitates the computation of HFGT mathematical models and provides a Graphical User Interface based Petri net simulator to visualize the mathematical models.  It is written in the MATLAB language and has been tested with v9.6 (R2019a). It is openly available on GitHub with a sample input file for straightforward re-use.

\section*{Keywords}

Hetero-functional Graph Theory; Model Based Systems Engineering; Petri nets 

\section*{Introduction}

In the \nth{20} century, newly invented technical artifacts products were connected to form large-scale complex engineering systems.   Indeed, the electric generator, the telephone, the petro-chemical refinery, the automobile and a whole host of medical treatment and imaging devices have given rise to the electric power, communication, oil \& gas pipeline, transportation, and healthcare delivery systems that we know today \cite{De-Weck:2011:00}.  Over time, these engineering systems have evolved to incorporate newer technologies that rely on multiple engineering systems (e.g. renewable energy, mobile phones, fuel-cells, electric-vehicles, and wearable-health technologies).  Consequently,  the interactions found within these networked systems have grown in both degree as well as heterogeneity.   Furthermore, these already complex engineering systems have \emph{converged} in what is now called \textit{systems-of-systems}. The \say{smart} grid, the energy-water nexus, electrified transportation systems, the energy-water-food nexus, and the development of interdependent smart-city infrastructure are all examples of how contemporary systems-of-systems are becoming an integral part of human life \cite{Farid:2016:ISC-BC06}. This paper regards such systems as engineering systems:

\begin{defn}
Engineering System \cite{De-Weck:2011:00} : A class of systems characterized by a high degree of technical complexity, social intricacy, and elaborate processes aimed at fulfilling important functions in society.
\end{defn}

\noindent The analysis, design, planning, and operation of these engineering systems from a holistic perspective has necessitated ever-more sophisticated modeling techniques. Two informatic sciences are of particular relevance. \textbf{\emph{Model-Based Systems Engineering}} (MBSE) is a practical and interdisciplinary engineering field that enables a successful realization of complex systems from concept, through design, to full implementation\cite{SE-Handbook-Working-Group:2015:00}.  Often, the practice of MBSE relies on the use of UML\cite{Rumbaugh:2005:00} and/or SysML\cite{Weilkiens:2007:00,Friedenthal:2011:00} which consists of several \emph{graphical} viewpoints of the function and form of the engineering system\cite{Crawley:2015:00}. These viewpoints include block diagrams, activity diagrams, and state-machine diagrams; making them well-equipped to describe the complex and heterogeneous nature of these systems; both graphically and later in simulation.  In the meantime, \textbf{\emph{network science}} has matured as a discipline to provide quantitative analyses for the structure and function of networks that appear across natural, social and engineering sciences\cite{Newman:2006:00,Barabasi:2016:00}. The spatially-distributed nature of engineering systems has often times led to their underlying models being rooted in graph theory.  Thus, it has been applied to systems like transportation systems\cite{XieF:2009:00}, power grids\cite{Albert:2004:00,Amaral:2000:00}, water networks\cite{DiNardo:2013:00}, supply chains\cite{Ivanov:2012:00}, and healthcare systems\cite{Yih:2016:00}.\\

\noindent Despite significant advancements, these seemingly disparate fields have experienced similar limitations in addressing the inherent complexity of engineering systems.  While the graphical models used in MBSE often serve as the basis for developing complex simulations of \emph{system behavior}, they fall short in providing a quantitative analysis of \emph{system structure}.  On the other hand, network science's reliance on graphs as a data structure limits its ability to handle the explicit heterogeneity one often encounters in engineering systems.  Even the recent developments toward multi-layer networks have been shown to exhibit several modeling constraints that inhibit the representation of an arbitrary number of network layers of arbitrary topology connected arbitrarily\cite{Kivela:2014:00,Schoonenberg:2018:ISC-BK04}.  The trend toward greater and more heterogeneous interaction between multiple engineering systems is only set to accelerate given the continual proliferation of diverse physical and information technologies\cite{Muhanji:2019:SPG-BK05,McFarlane:2003:00,Mcfarlane:2002:06,McFarlane:2013:00,Meyer:2009:00}.  The current methodological and theoretical limitations in the MBSE and network science fields necessitates new mathematical modeling techniques for multiple  integrated engineering systems.  \\

\noindent Hetero-Functional Graph Theory (HFGT) has emerged as a means to model the structure and function of highly interconnected and heterogeneous engineering systems \cite{Schoonenberg:2018:ISC-BK04}. In order to support insightful quantitative analyses, HFGT relies on multiple graphs as data structures. It also explicitly incorporates the heterogeneity of conceptual and ontological constructs found in MBSE. In doing so, it facilitates the translation of systems engineering models (e.g. SysML) to a mathematical engineering systems description, providing a rigorous platform for modeling systems-of-systems. The foundational HFGT works are in the field of mass-customized production systems \cite{Farid:2006:IEM-C02,Farid:2007:IEM-TP00,Farid:2008:IEM-J04,Farid:2008:IEM-J05}. In some ways, such production systems present modelling challenges that are common to most engineering systems. The production capabilities of a production system can come together in various permutations and combinations to produce an almost infinite number of product variants. At the same time, their structure and behavior is dynamically changing. Since these first works, the theory has methodologically evolved to provide qualitative and quantitative analyses in other application domains including electric power systems\cite{Farid:2015:SPG-J17,Thompson:2020:SPG-C68}, energy-water nexus \cite{Lubega:2014:EWN-J11,Carey:2018:EWN-P04,Thompson:2019:EWN-W10,Schoonenberg:2018:ISC-BK04}, electrified transportation systems\cite{Farid:2015:ETS-J25,Farid:2016:ETS-J27,vanderWardt:2017:ETS-J33}, microgrid-enabled production systems\cite{Schoonenberg:2017:IEM-J34}, industrial energy management\cite{Schoonenberg:2015:IEM-C48}, personalized healthcare delivery systems\cite{Khayal:2017:ISC-J35,Khayal:2016:ISC-J31,Khayal:2020:ISC-JR01} and interdependent smart city infrastructures\cite{Schoonenberg:2018:ISC-BK04}.

\noindent Hetero-functional graph theory is composed of seven mathematical models that together form a System Adjacency Matrix $\mathbb{A}$ as its eighth.
\begin{enumerate}
    \vspace{-0.05in}
    \item System Concept  $A_S$
    \vspace{-0.05in}
    \item Hetero-functional Adjacency Matrix $A_\rho$
    \vspace{-0.05in}
    \item Hetero-functional Incidence Tensor ${\cal M}_\rho$
    \vspace{-0.05in}
    \item Controller Agency Matrix $A_Q$
    \vspace{-0.05in}
    \item Controller Adjacency Matrix $A_C$
    \vspace{-0.05in}
    \item Service as Operand Behavior $N_L$
    \vspace{-0.05in}
    \item Service Feasibility Matrix $\Lambda$
    \vspace{-0.05in}
\end{enumerate}
\noindent This paper serves as a user guide to a recently developed \emph{Hetero-functional Graph Theory Toolbox}\cite{Hegde:2020:ISC-W11} which facilitates the instantiation of these seven mathematical models from a single XML input file.  It is written in the MATLAB\cite{MATLAB:2019} language and has been tested with v9.6 (R2019a). It is openly available on GitHub\cite{Hegde:2020:ISC-W11} together with a sample input XML file for straightforward re-use.  Additionally this paper serves as a user guide to a Graphical User Interface (GUI) that facilitates the use of the HFGT toolbox and an event-driven Comma Separated Value (CSV) file to visualize flows of operands through the HFG.\\

\noindent The remainder of the paper proceeds as follows.  The section entitled:  \say{Creating a HFGT Input File} details how to translate a real-world system into an HFGT Toolbox input XML file.  The following section entitled \say{HFGT Data Structures and Functions} describes the {\tt myLFES} (LFES: Large Flexible Engineering System) data structure which organizes all of the data generated by the toolbox.  This section also describes the functions that transform the input XML file into an empty version of the {\tt myLFES} data structure.  Lastly, it describes the functions that populate the myLFES data structure with HFGT values.  The section entitled \say{Toolbox validation} describes how the HFGT toolbox has been validated against previously published results.The next section entitled \say{Petri net Visualization} describes the additional input information required for visualizing the HFG in a Petri net.  This section also describes how the {\tt myPetriNetwork} system structure is structured.  Finally, the \say{Conclusions} section brings the paper to a close.  This paper presumes that the reader has a working knowledge of HFGT which is otherwise gained from a thorough reading of the associated text \cite{Schoonenberg:2018:ISC-BK04}.  It is also assumed that the reader has a working knowledge of Petri nets which is otherwise gained from the associated text \cite{Murata:1989:01,Jensen:1992:01,Girault:2013:00}.  Furthermore, the HFGT toolbox uses an XML input file and so a basic knowledge of the eXtensible Markup Language\cite{Elliotte:2002:00} is needed.  Additionally, the Tensor Toolbox for MATLAB \cite{TTB_Sparse, TTB_Software} is used to store incidence tensors.  Finally, UML/SysML models\cite{Weilkiens:2007:00} are used to convey the object-oriented programming data structures found throughout the HFGT toolbox.  

\section*{Implementation and architecture}

\section*{Creating a HFGT XML Input File}
\begin{figure}[!ht]
\centering
\vspace{-0.2in}
\includegraphics[width=0.9\linewidth]{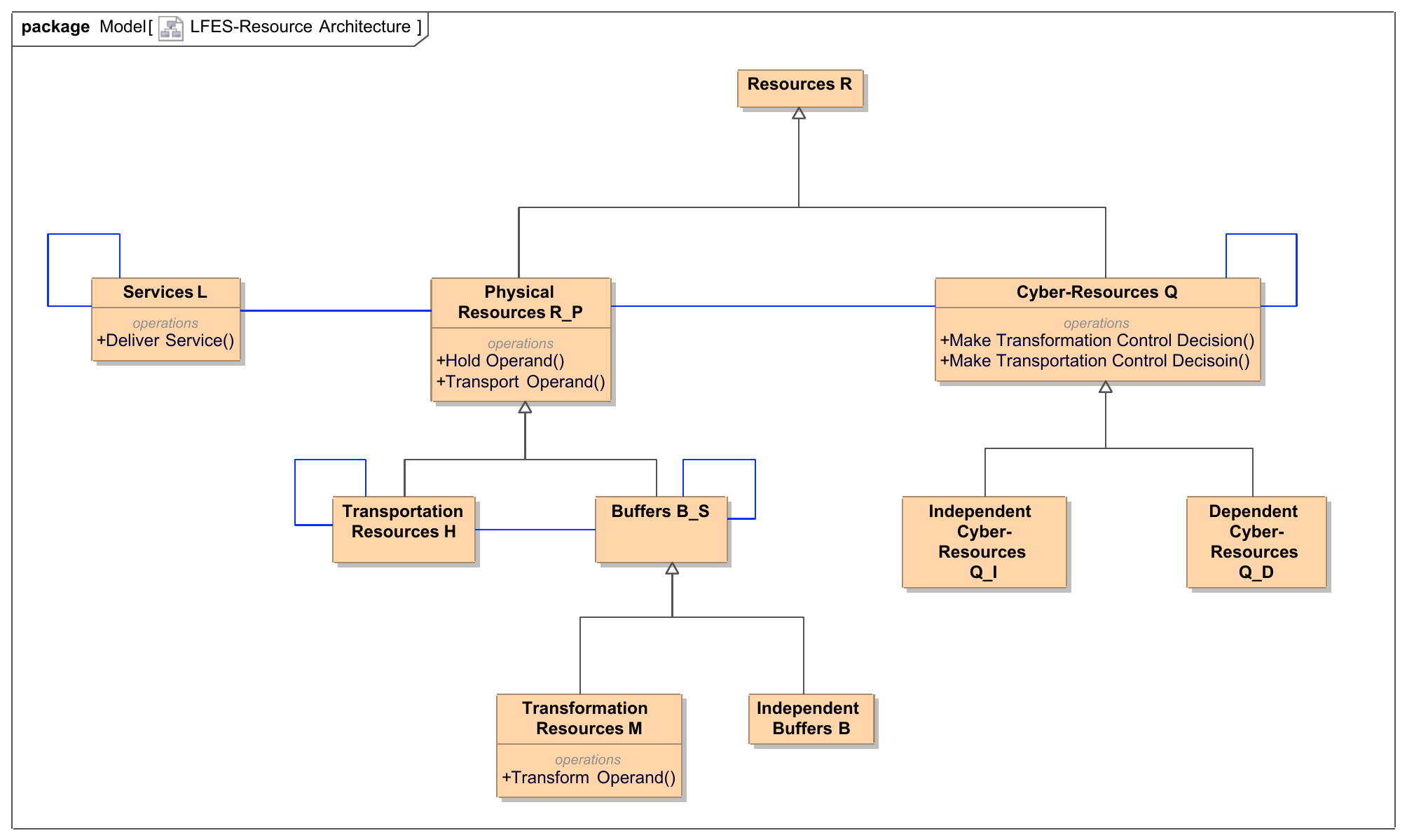}
\caption{A SysML Block Diagram of the System Form of the LFES Meta-Architecture}
\label{fig:LFES_ResourceArchitecture}
\vspace{-0.1in}
\end{figure}

\begin{figure}[!ht]
\centering
\includegraphics[width=\linewidth]{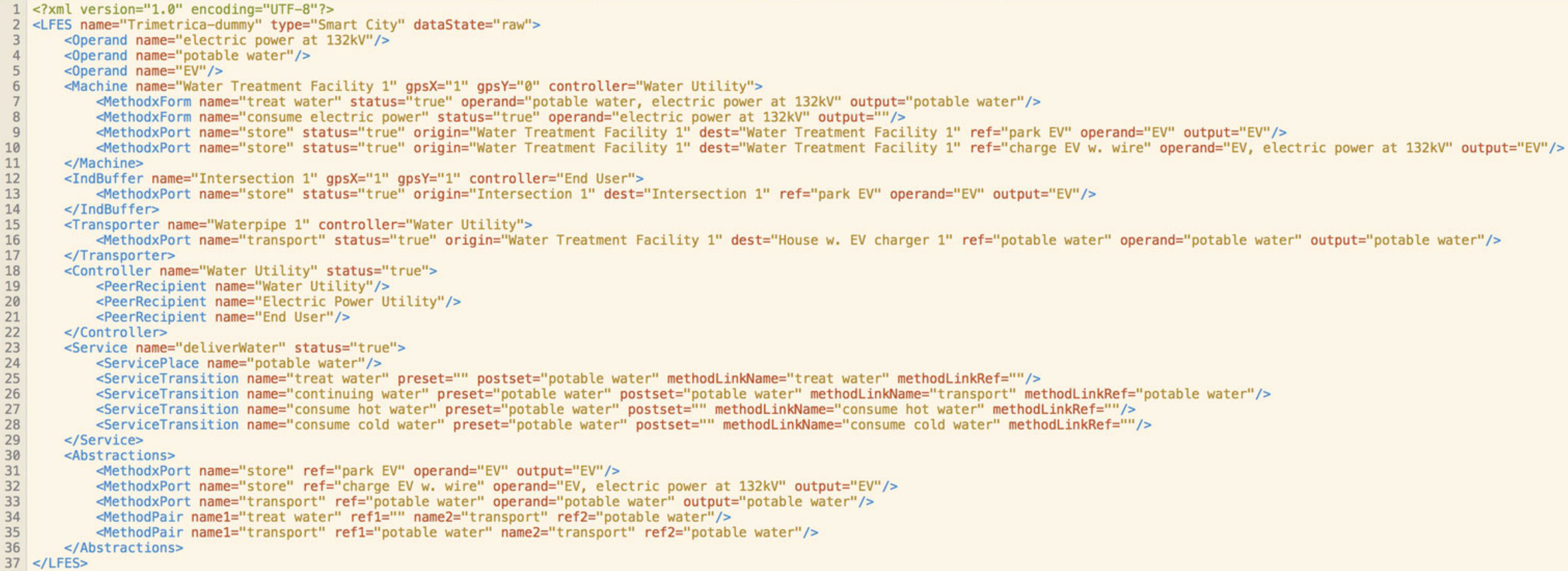}
\caption{An example HFGT Toolbox input XML input file}
\label{fig:Trimetrica_Dummy}
\vspace{-0.15in}
\end{figure}

\noindent As a high-level overview, the purpose of the HFGT input file is to provide a structured representation of the data associated with an instantiated large flexible engineering systems.  To that end, Figure \ref{fig:LFES_ResourceArchitecture} shows the meta-architecture of the system form of a Large Flexible Engineering System (LFES) as described by HFGT\cite{Schoonenberg:2018:ISC-BK04}. HFGT assumes that the formal elements of any LFES can be viewed as instances of this meta-architecture.  The input data file, thus, organizes the resources present in an engineering system into these \say{meta-elements} and captures the functions (i.e. methods) that they are capable of performing.  Such a complex hierarchical structure is readily stated in the eXtensible Markup Language (XML)\cite{Elliotte:2002:00}.  Furthermore, XML files are both human and machine readable; making them easy to check for inadvertently introduced errors. \\

\noindent The first step to using the HFGT toolbox is to accurately write an input XML file that instantiates the meta-architecture shown in Figure \ref{fig:LFES_ResourceArchitecture}.  Figure \ref{fig:Trimetrica_Dummy} provides an example of such an XML file.  Each line of the XML file is composed of a start or end XML element (e.g <LFES>) that corresponds to one of the classes identified in Figure \ref{fig:LFES_ResourceArchitecture}.  Furthermore, each of these elements has one or more attributes that correspond to the methods and attributes of these classes.  Each of these XML elements are now explained in turn.  

\begin{trivlist}
\item \textbf{LFES}:  This element is the root of the input XML file and indicates an instantiated LFES.  All of the meta-elements shown in Figure \ref{fig:LFES_ResourceArchitecture}, and the processes that they are capable of carrying out are contained within this root. The attributes “name” and “type” describe the name of the system and its type. The attribute “dataState” is defaulted to “raw” to indicate its pre-processed state.  
\item \textbf{Machine}: This XML element is representative of a “Transformation Resource” (M) shown in Figure \ref{fig:LFES_ResourceArchitecture}. The “name” attribute describes the name of the transformation resource. If a controller has agency over the transformation resource, the former's name is stored under the “controller” attribute. The attributes “gpsX” and “gpsY” capture the physical location of the transformation resource with respect to a reference axes. The illustrative example in Figure \ref{fig:Trimetrica_Dummy} shows a water treatment facility as an instantiaton of a transformation resource.
\item \textbf{MethodxForm}: This XML element is used to describe the ability of a transformation resource to transform an operand. As shown in Figure \ref{fig:LFES_ResourceArchitecture} and as reflected in the illustrative example in Figure \ref{fig:Trimetrica_Dummy}, this element is nested within the Machine (i.e. transformation resources) element. The attributes “name”, “status”, “operand”, and “output” describe the name of the process, its status (active/inactive), the set of operands needed for the process, and the set of outputs as a result of the transformation respectively.
\item \textbf{MethodxPort}:  As shown in Figure \ref{fig:LFES_ResourceArchitecture}, the operations involving holding or transportation of an operand are universal to all of the physical resources. Rather than introduce an XML element for holding and transportation processes separately, the XML file format introduces this element to refer to a ``refined transportation process"\cite{Schoonenberg:2018:ISC-BK04} as their combination.  Furthermore, a storage process is considered a transportation with the same origin and destination.  The attributes used to describe these process include “name”, “status”, "origin", and "dest", which describe the name, the active state, the origin and the destination of the refined transportation process respectively. The attribute \say{ref} describes the refinement of the transportation in terms of a string that reflects the associated holding process.  The operands needed to carry out the process and the output of the process are described by the attributes “operand” and “output” respectively. 
\item \textbf{IndBuffer}: This XML element is representative of an “Independent Buffer” (B) as shown in Figure \ref{fig:LFES_ResourceArchitecture}. The “name” attribute describes the name of the independent buffer.  If a controller has agency over the independent buffer, the former's name is stored under the “controller” attribute. The attributes “gpsX” and “gpsY” capture the physical location of the independent buffer with respect to a reference axes. 
\item \textbf{Transporter}: This XML element represents a “Transportation Resource” (H) as shown in Figure \ref{fig:LFES_ResourceArchitecture}. The attributes “name” and “controller” captures its name and the controller that controls the resource respectively.
\item \textbf{Controller}: As shown in Figure \ref{fig:LFES_ResourceArchitecture}, controllers have jurisdiction or agency over physical resources. The association of a physical resource to an independent controller is captured in the attributes of the physical resource itself. Dependent controllers embedded within a physical resource are implicit to the tooolbox's functionality and are not explicitly stated in the input XML File.  Thus, the \say{controller} XML element describes independent controllers exclusively. The attribute \say{name} describes the name of the controller and the attribute \say{status} describes whether the controller is active or inactive.  
\item \textbf{PeerRecipient}: This XML element describes the peer controllers that \emph{receive information from} the controller named in the controller XML element.  The name attribute provides the name of this peer controller.  
\item \textbf{Service}: The services shown in Figure \ref{fig:LFES_ResourceArchitecture} describe the evolution in the state of an operand as a delivered service modeled as a Petri Net\cite{Murata:1996:00}.  The attribute \say{name} describes the name of the service and the attribute \say{status} describes whether the service is active or inactive in the system.  
\item \textbf{ServicePlace}: This XML element describes the places of the service (Petri) net.  Each place is given its own name under the ``name" attribute.  
\item \textbf{ServiceTransition}: This XML element describes the transitions of the service (Petri) net.  Each transition is given its own name under ``name" attribute.  The ``preset" and ``postset" attributes refer to the respective names of the places that send tokens to or receive tokens from the service transition named in this XML element.  The ``methodLinkName" attribute indicates the name of the system process to which the transition is linked.  The ``methodLinkRef" attribute indicates the refinement of the transportation process to which the transition is linked (if any).  As shown in Figure \ref{fig:LFES_ResourceArchitecture}, services and physical resources are associated by virtue of the link between the service transitions in the former and the system processes provided by the latter.    
\item \textbf{Abstractions}: This XML element has no attributes associated with it. It contains all possible holding refinements and functional sequences (i.e. MethodPairs) in the system nested within it.
\item \textbf{MethodxPort} (within the Abstractions Element): This XML element contains information about all possible refinements and their associated operands and outputs in the system. In the illustrative example in Figure \ref{fig:Trimetrica_Dummy}, the \say{refinements} possible while transporting the Electric Vehicle (EV) are: parking the EV, charging it by wire, discharging it and charging it wirelessly.
\item \textbf{MethodPair} (within the Abstractions Element): This XML element captures the possibility of one system process following another in the system. In Figure \ref{fig:Trimetrica_Dummy}, the possibility that water being transported from an origin to a destination can be followed by a transportation from the new origin to a new destination has been described. 
\end{trivlist}

\section*{HFGT Data Structures and Functions}
\begin{figure}[!ht]
\centering
\vspace{-0.1in}
\includegraphics[width=0.78\linewidth]{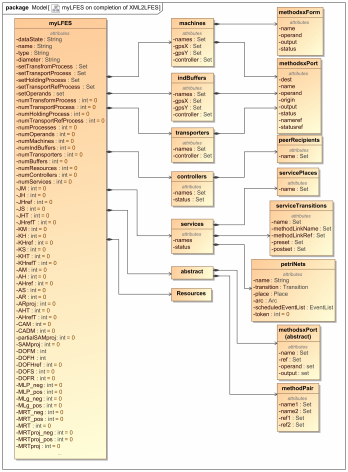}
\caption{Structure of myLFES on completion of XML2LFES}
\label{fig:myLFESafterXML2LFES}
\vspace{-0.2in}
\end{figure}

\begin{figure}[!ht]
\centering
\includegraphics[width=0.95\linewidth]{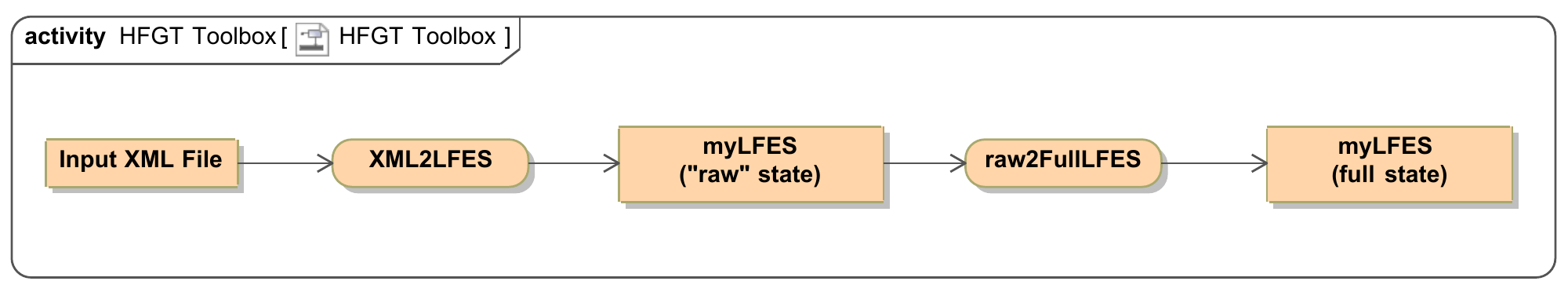}
\caption{A high-level activity diagram of the HFGT Toolbox}
\label{fig:HFGT_Toolbox_overview}
\vspace{-0.15in}
\end{figure}

The \emph{Hetero-functional Graph Theory Toolbox} uses MATLAB's object-oriented programming functionality to enhance modularity, extensibility and reusability.   The primary purpose of the HFGT toolbox is to instantiate and subsequently populate the data in the {\tt myLFES} structure.  It's associated class diagram is shown in Figure \ref{fig:myLFESafterXML2LFES}.  The {\tt myLFES} stores all of the information in the XML input file and then calculates all of the mathematical quantities identified in HFGT.     
\noindent This high-level purpose is achieved through two principal modules: {\tt XML2LFES()} and {\tt raw2FullLFES()} that are executed in sequence as shown in Figure \ref{fig:HFGT_Toolbox_overview}.  In brief, the {\tt XML2LFES()} module serves to import the input XML file and create the {\tt myLFES} data structure in a ``raw" structure.  Then, the {\tt raw2Full()} module makes the HFGT calculations necessary to convert the {\tt myLFES} data structure to the ``full" state.  Each of these modules is now discussed in detail.  

\vspace{0.2in}
\subsection*{[myLFES,S] = XML2LFES(XMLFile)}
\begin{figure}[!ht]
\centering
\includegraphics[width=0.95\linewidth]{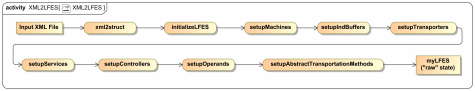}
\caption{Activity diagram of the XML2LFES module}
\label{fig:XML2LFES}
\vspace{-0.15in}
\end{figure}

As previously mentioned, the {\tt XML2LFES()} module serves to import the input XML file and create the {\tt myLFES} data structure in a ``raw" structure.  This functionality is achieved through the sequence of functions shown in the activity diagram in Figure \ref{fig:XML2LFES}.  Each of these functions are now explained in turn.  



\begin{trivlist}
\item \textbf{S  = xml2struct(xmlfile)} : The {\tt xml2struct} function (v1.8.0.0) is a freely available third-party MATLAB function that can be otherwise downloaded from the MATLAB Central website\cite{Falkena:2020:00}.  In the context of the HFGT toolbox, this function reads the input XML file and converts it into an intermediate object {\tt S} which replicates the hierarchy of the input XML file. The {\tt S} structure is identical to the {\tt myLFES} structure shown in Figure \ref{fig:myLFESafterXML2LFES} with two exceptions:  
\end{trivlist}
\begin{enumerate}
\item It does not contain the root attributes of myLFES.  These are calculated later.  
\item At the lowest level of decomposition, {\tt S} has a cell array of objects (e.g. {\tt S.myLFES.machines\{:\}.methodsxForm}).  In {\tt myLFES}, the lowest level of decomposition has an object of cell arrays (e.g. {\tt myLFES.machines...} \\ {\tt .methodsxForm\{:\}}).  
\end{enumerate}
Because the {\tt xml2struct()} function is relatively rigid, and there is not much control on its output {\tt S}, the remaining functions in {\tt XML2LFES()} serve to ``reshape" {\tt S} into the more desirable form {\tt myLFES} without any loss of information.  
\begin{trivlist}
\item \textbf{myLFES = initializeLFES()} : This function is the LFES constructor and consequently instantiates {\tt myLFES}.  The high-level integer attributes of {\tt myLFES} are initialized to zero and the sub-classes are initialized as empty structured arrays.  It also calculates the number of transformation and structural degrees of freedom in {\tt myLFES.DOFM} and {\tt myLFES.DOFH} respectively.  

\item \textbf{myLFES = setupMachines(myLFES, S)} : This function assigns information to the sub-object {\tt myLFES.machines} (M) by \say{unpacking} {\tt S.myLFES.machines}.   It also adds the {\tt nameref} and {\tt statusref} attributes to the {\tt myLFES.machines...} {\tt.methodsxPort} ($P_\eta$) class.  These refer to the name of the associated transportation process and its initial status respectively.  This function also collates the set of all transformation processes ($P_\mu$) in {\tt myLFES.setTransformProcess}.  

\item \textbf{myLFES = setupIndBuffers(myLFES, S)} :This function assigns information to the sub-object {\tt myLFES.indBuffers} (B) by \say{unpacking} 
{\tt S.myLFES.indBuffers}.  It also adds the {\tt nameref} and {\tt statusref} attributes to the {\tt myLFES.} {\tt...indBuffers.methodsxPort} ($P_\eta$) class.  These refer to the name of the associated transportation process and its initial status respectively.  It also increments the value of {\tt myLFES.DOFH} with the capabilities of the independent buffers.  

\item \textbf{myLFES = setupTransporters(myLFES, S)} : This function assigns information to the sub-object {\tt myLFES...}\\ {\tt .transporters} (H) by \say{unpacking} {\tt S.myLFES.transporters}.  It also adds the {\tt nameref} and {\tt statusref} attributes to the {\tt myLFES...} {transporters.methodsxPort} ($P_\eta$) class.  These refer to the name of the associated transportation process and its initial status respectively.  It also increments the value of {\tt myLFES.DOFH} with the capabilities of the transportation resources. 

\item \textbf{myLFES = setupServices(myLFES, S)} : This function assigns information to the sub-object {\tt myLFES.services} by \say{unpacking} 
{\tt S.myLFES.services}.    

\item \textbf{myLFES = setupControllers(myLFES, S)} : This function assigns information to the sub-object {\tt myLFES.controllers} (Q) by \say{unpacking} 
{\tt S.myLFES.controllers}.

\item \textbf{myLFES = setupOperands(myLFES, S)} : This function assigns information to the set {\tt myLFES.setOperands} (L) and {\tt myLFES.numOperands} by \say{unpacking} {\tt S.myLFES.Operand}.

\item \textbf{setupAbstractTransportationMethods(myLFES, S)} : This function assigns information to the sub-object {\tt myLFES.} \\ {\tt..abstract} by \say{unpacking} {\tt S.myLFES.abstract}.
\end{trivlist}

\noindent The following support functions are used within the functions that set up the elements of the system: 
\begin{trivlist}
    \item \textbf{objB=getResourceAttributes(objA)} : This function copies the attributes of a resource from the intermediate structure {\tt S} to the corresponding resource class object within the {\tt myLFES} object. This is helpful in specifying the attributes of each type of meta-element within {\tt myLFES}.
    \item \textbf{[objB, setProcess, DOF]=getResourceMethods(objA,setProcess,DOF,opt)} : This function copies information associated with a resource from the intermediate structure {\tt S} to the corresponding resource class object within the {\tt myLFES} object. The {\tt opt} argument of the function follows the classification in Figure 2. It is assigned \say{M},\say{B} or \say{H} for machines, independent buffers and transporters respectively. In addition, it computes the transportation and transportation capabilities. 
    
    \item \textbf{objA=insertObjB(objA,objB,idxA)} : This function inserts the fields of {\tt objB} into the equivalent fields of {\tt objA}. This function is used as a support function within both {\tt getResourceAttributes} and {\tt getResourceMethods} to extract fields from structure {\tt S} and fit it into the correct hierarchical location within {\tt myLFES}. 
\end{trivlist}

\subsection*{myLFES=raw2FullLFES(myLFES)} 
\begin{figure}[!ht]
\centering
\includegraphics[width=0.95\linewidth]{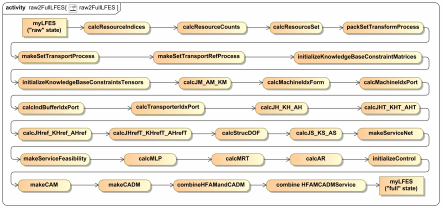}
\caption{Activity diagram of the raw2Full module}
\label{fig:raw2Full}
\vspace{-0.15in}
\end{figure}

As mentioned previously, the {\tt raw2FullLFES()} module makes the HFGT calculations necessary to convert the {\tt myLFES} data structure from the ``raw" to the ``full" state.  This functionality is achieved through the sequence of functions shown in the activity diagram in Figure \ref{fig:raw2Full}.  Each of these functions are now explained in turn.  

\begin{trivlist}
\item \textbf{myLFES=calcResourceIndices(myLFES)} : This function assigns the new attributes {\tt idxMachine}, {\tt idxBuffer} and \newline {\tt idxTransporter} to the sub-objects {\tt myLFES.machines} (M), {\tt myLFES.indBuffers} (B), and {\tt myLFES.transporters} (H) respectively. These attributes represent the numeric index of the resource for each type of physical resource in the system. In addition, it also assigns an attribute {\tt idxResource} to the sub-objects {\tt myLFES.machines} (M), {\tt myLFES.indBuffers} (B), and {\tt myLFES.transporters} (H) which indicates the index of the resource in the set of resources.      

\item \textbf{myLFES=calcResourceCounts(myLFES)} : This function calculates the number of each type of resource present in the system. In doing so, it populates the following attributes:  {\tt myLFES.numMachines}, {\tt myLFES.numIndBuffers}, \newline {\tt myLFES.numBuffers}, {\tt myLFES.numTransporters}, {\tt myLFES.numResources}, {\tt myLFES.numControllers} and {\tt myLFES.} \\ {\tt ...numServices}.

\item \textbf{myLFES=calcResourceSet(myLFES)} : This function creates a new class object {\tt myLFES.resources} with two attributes:  {\tt names} and {\tt idx}.  These are the set unions of the {\tt names} and {\tt idxResource} attributes in {\tt myLFES.machines} (M), \newline {\tt myLFES.indBuffers} (B) and {\tt myLFES.transporters} (H).

\item \textbf{myLFES=packSetTransformProcess(myLFES)} : This function computes the set of unique transformation processes and assigns it to  {\tt myLFES.setTransformProcess} ($P_\mu$). It also assigns the number of unique transformation processes to {\tt myLFES.numTransformProcess}.

\item \textbf{myLFES=makeSetTransportProcess(myLFES)} : This function computes the set of all possible transportation processes system and assigns it to {\tt myLFES.setTransportProcess} ($P_\eta$). It also assigns the number of transportation processes to {\tt myLFES.numTransportProcess}.

\item \textbf{myLFES=makeSetTransportRefProcess(myLFES)} : This function computes the set of all possible refined transportation processes and assigns it to {\tt myLFES.setRefTransportProcess} ($P_{\bar{\eta}}$). It also uses the information in {\tt myLFES.abstract$\{p_{\gamma g}\}$} \newline {\tt ...methodsxport} to compute the set of holding processes and assign the result to {\tt myLFES.setHoldingProcess} ($P_\gamma$). It also computes the numbers of refined transportation processes and holding processes and assigns them to \newline {\tt myLFES.numTransportProcess} and {\tt myLFES.numHoldingProcess} respectively.

\item \textbf{myLFES=initializeKnowledgeBasesConstraintsMatrices(myLFES)} : This function creates appropriately sized but empty sparse matrices to the three knowledge bases (i.e. {\tt myLFES.JM} ($J_M$), {\tt myLFES.JH} ($J_H$), {\tt myLFES.JHRef}) ($J_{\bar{H}}$), the three constraint matrices (i.e. {\tt myLFES.KM} ($K_M$), {\tt myLFES.KH} ($K_H$), {\tt myLFES.KHRef}) ($K_{\bar{H}}$), and the three system concept matrices (i.e. {\tt myLFES.AM} ($A_M$), {\tt myLFES.AH} ($A_H$), {\tt myLFES.AHRef} ($A_{\bar{H}}$)). 

\item \textbf{myLFES=initializeKnowledgeBasesConstraintsTensors(myLFES)} : This function creates appropriately sized but empty sparse tensors to two knowledge bases (i.e. {\tt myLFES.JHT} (${\cal J}_H$), {\tt myLFES.JHRefT} (${\cal J}_{\bar{H}}$)), the two constraint tensors (i.e. {\tt myLFES.KHT} (${\cal K}_H$), {\tt myLFES.KHRefT} (${\cal K}_{\bar{H}}$)), and the two system concept tensors (i.e. {\tt myLFES.AHT} (${\cal A}_H$), {\tt myLFES.AHRefT} (${\cal A}_{\bar{H}}$)).

\item \textbf{myLFES=calcJM\_KM\_AM(myLFES)} : This function computes the transformation knowledge base ({\tt myLFES.JM}, $J_M$), the transformation constraint matrix ({\tt myLFES.KM}, $K_M$) and the resultant transformation system concept ({\tt myLFES.AM}, $A_M$).   

\item \textbf{myLFES=calcMachineIdxPort(myLFES)} :  This function computes the index associated with the transformation process conducted by a given machine.  It thus populates {\tt myLFES.machines.methodsxForm.idxForm}, the index of the transformation process from the set of transformation processes possible in the system.

\item \textbf{myLFES=calcMachineIdxPort(myLFES)} :  This function computes the following five indices associated with a refined transportation process conducted by a given machine.  
\end{trivlist}
\begin{itemize}
\item {\tt myLFES.machines.methodsxPort.idxOrigin} : the index of the resource where the transportation/holding process originates from the list of indices of all resources present in the system.

\item {\tt myLFES.machines.methodsxPort.idxDest} : the index of the resource where the transportation/holding process terminates from the list of indices of all resources present in the system.

\item {\tt myLFES.machines.methodsxPort.idxHold} : the index of the holding process from the list of holding processes in the system. 

\item {\tt myLFES.machines.methodsxPort.idxPort} : the index of the transportation process from the set of transportation processes possible in the system.

\item {\tt myLFES.machines.methodsxPort.idxPortRef} : the index of the refined transportation process from the set of refined transportation processes possible in the system.
\end{itemize}
\begin{trivlist}
\item \textbf{myLFES=calcIndBufferIdxPort(myLFES)}:  This function is analogous to {\tt calcMachineIdxPort()} above, but instead calculates the index attributes in {\tt myLFES.indBuffers.methodsxPort}.  

\item \textbf{myLFES=calcTransporterIdxPort(myLFES)}:  This function is analogous to {\tt calcMachineIdxPort()} above, but instead calculates the index attributes in {\tt myLFES.transporters.methodsxPort}.  

\item \textbf{myLFES=calcJH\_KH\_AH(myLFES)} : This function computes the transportation knowledge base matrix ({\tt myLFES.JH}, $J_H$), the transportation constraint matrix ({\tt myLFES.KH}, $K_H$) and the transportation system concept matrix ({\tt myLFES.AH}, $A_H$).  

\item \textbf{myLFES=calcJHT\_KHT\_AHT(myLFES)} : This function computes the transportation knowledge base tensor ({\tt myLFES.JHT}, ${\cal J}_H$), the transportation constraint tensor ({\tt myLFES.KHT}, ${\cal K}_H$) and the transportation system concept tensor ({\tt myLFES.AHT}, ${\cal A}_H$).

\item \textbf{myLFES=calcJHref\_KHref\_AHref(myLFES)} : This function computes the refined transportation knowledge base \newline ({\tt myLFES.JHref}, $J_{\bar{H}}$), the refined transportation constraint matrix ({\tt myLFES.KHref}, $K_{\bar{H}}$) and the refined transformation system concept ({\tt myLFES.AHref}, $A_{\bar{H}}$).

\item \textbf{myLFES=calcJHrefT\_KHrefT\_AHrefT(myLFES)} : This function computes the refined transportation knowledge base tensor ({\tt myLFES.JHrefT}, ${\cal J}_{\bar{H}}$), the refined transportation constraint tensor ({\tt myLFES.KHrefT}, ${\cal K}_{\bar{H}}$) and the refined transportation system concept tensor ({\tt myLFES.AHrefT}, ${\cal A}_{\bar{H}}$).

\item \textbf{myLFES=calcStrucDOF(myLFES)} : This function calculates the number of independent actions that completely define the number of available transformation, transportation and refined transportation processes in a system and assigns it to {\tt myLFES.DOFM}, {\tt myLFES.DOFH}, {\tt myLFES.DOFHref} respectively.

\item \textbf{myLFES=calcJS\_KS\_AS(myLFES)} : This function computes the system knowledge base ({\tt myLFES.JS}, $J_S$), the system constraint matrix ({\tt myLFES.KS}, $K_S$) and the system concept (\textbf{{\tt myLFES.AS}}, $A_S$). 

\item \textbf{myLFES=makeServiceNets(myLFES)} : This function captures information regarding the different states involved in delivering an operand in the system and translates it into a service Petri net ($N_{l_i}$) with places ($S_{l_i}$), transitions (${\cal E}_{l_i}$), and arcs ($M_{l_i}$) that defines the adjacency of the service activities in the system. For each service, it stores the positive incidence matrix, the negative incidence matrix and the dual-adjacency matrix under {\tt myLFES.services.Mpos} ($M_{l_i}^+$), {\tt myLFES.services.Mneg} ($M_{l_i}^-$) and {\tt myLFES.services.dualAdjacency} ($A_{l_i}$) respectively.

\item \textbf{myLFES=makeServiceFeasibility(myLFES)} : This function computes the service feasibility matrix. It uses the structural degrees of freedom previously computed in the toolbox to compute the service degrees of freedom. It computes and assigns the following attributes to the {\tt myLFES} object:
\end{trivlist}
\begin{itemize}
    \item {\tt myLFES.services.rawLambda} ($\Lambda_i$): Service feasibility matrix in its original shape for each deliverable service.
    \item {\tt myLFES.services.rawLambda\_neg} ($\Lambda_i^-$): Negative component of the Service feasibility matrix in its original shape for each deliverable service.
    \item {\tt myLFES.services.rawLambda\_pos} ($\Lambda_i^+$): Positive component of the Service feasibility matrix in its original shape for each deliverable service.
    \item {\tt myLFES.services.Lambda} ($\Lambda_{S_i}$): Service feasibility matrix projected to DOFs for each deliverable service.
    \item {\tt myLFES.services.RawxFormLambda} ($\Lambda_{\mu i}$): Service transformation feasibility matrix in its original shape for each deliverable service.
    \item {\tt myLFES.services.RawxFormLambda\_neg} ($\Lambda_{\mu i}^-$): Negative Service transformation feasibility matrix in its original shape for each deliverable service.
    \item {\tt myLFES.services.RawxFormLambda\_pos} ($\Lambda_{\mu i}^+$): Positive Service transformation feasibility matrix in its original shape for each deliverable service.
    \item {\tt myLFES.services.xFormLambda} ($\Lambda_{Mi}$): Service transformation feasibility matrix projected to system capabilities for each deliverable service.
    \item {\tt myLFES.services.xPortLambda} ($\Lambda_{\gamma i}$): Service transportation feasibility matrix for each deliverable service.
    \item {\tt myLFES.services.xPortLambda\_neg} ($\Lambda_{\gamma i}^-$): Negative Service transportation feasibility matrix for each deliverable service.
    \item {\tt myLFES.services.xPortLambda\_pos} ($\Lambda_{\gamma i}^+$): Positive Service transportation feasibility matrix for each deliverable service.
\end{itemize}
\begin{trivlist}

\item \textbf{myLFES=calcMLP(myLFES)} : This function computes the positive and negative Holding Process-Operand Incidence Matrices \textbf{{\tt myLFES.MLg\_neg} ($M_{LP_\gamma}^-$), {\tt myLFES.MLg\_pos}} ($M_{LP_\gamma}^+$) and the positive and negative Process-Operand Incidence Matrices \textbf{{\tt myLFES.MLP\_neg} ($M_{LP}^-$), {\tt myLFES.MLP\_pos} ($M_{LP}^+$)}. 

\item \textbf{myLFES=calcMRT(myLFES)} : This function computes the positive and negative Hetero-functional Incidence Tensors \textbf{{\tt myLFES.MRT\_neg} (${\cal M}_{\rho}^-$), {\tt myLFES.MRT\_pos} (${\cal M}_{\rho}^+$), {\tt myLFES.MRT} (${\cal M}_{\rho}$)} and the positives and negative Projected Hetero-functional Incidence Tensors \textbf{{\tt myLFES.MRTproj\_neg} ($\tilde{{\cal M}}_{\rho}^-$), {\tt myLFES.MRTproj\_pos} ($\tilde{{\cal M}}_{\rho}^+$), {\tt myLFES.MRTproj}} ($\tilde{{\cal M}}_{\rho}$).

\item \textbf{myLFES=calcAR(myLFES)} : This function computes the Hetero-Functional Adjacency Matrix \textbf{{\tt myLFES.AR}} ($A_\rho$) and the projected Hetero-Functional Adjacency Matrix \textbf{{\tt myLFES.ARproj} ($\tilde{A}_{\rho}$)}. In parallel, the function calculates the number of physical continuity degrees of freedom and assigns their values to {\tt myLFES.DOFR1}, {\tt myLFES.DOFR2},{\tt  myLFES.}\\ {\tt ...DOFR3}, and {\tt myLFES.DOFR4} for Types 1 through 4 respectively. Furthermore, it computes the number of functional sequence dependent degrees of freedom and assigns it to {\tt myLFES.DOFR5}. Finally, the function computes the total degree of freedom sequences in the system and stores it as {\tt myLFES.DOFR}.

\item \textbf{myLFES = initializeControl(myLFES)} : This function creates appropriately sized but empty sparse matrices to the controller agency matrix ({\tt myLFES.CAM}, $A_Q$), the controller adjacency matrix ({\tt myLFES.CADM}, $A_C$) and the projected system adjacency matrix without services ({\tt myLFES.partialSAMproj} $\Tilde{\mathds{A}}$)

\item \textbf{myLFES=makeCAM(myLFES)} : This function computes the controller agency matrix and assigns it to {\tt myLFES.CAM} ($A_Q$).

\item \textbf{myLFES=makeCADM(myLFES)} : This function computes the controller adjacency matrix and assigns it to \newline {\tt myLFES.CADM} ($A_C$). 

\item \textbf{myLFES=combineHFAMandCADM(myLFES)} : This function computes the system adjacency matrix (without services) by combining the hetero-functional adjacency, the controller agency and the controller adjacency matrices.  It is assigned to {\tt myLFES.partialSAMproj} ($\Tilde{\mathds{A}}$).  

\item \textbf{myLFES=combineHFAMCADMService(myLFES)} : This function combines the hetero-functional adjacency, controller agency, controller adjacency, and service graph matrices to produce the system adjacency matrix. This combined matrix is assigned to {\tt myLFES.SAMproj} ($\Tilde{\mathds{A}}$).  
\end{trivlist}
\vspace{-0.25in}

\section*{Toolbox Validation} 
The accurate production of a hetero-functional graph (HFG) as an instantiated mathematical model is similar to that of the creation of a formal graph based exclusively upon nodes and edges. Formal graph toolboxes are able to create formal graphs as instantiated mathematical models because they reproduce the results that would be achieved for small systems computed by hand. In other words, these toolboxes serve to automate the construction of these mathematical models that would be impractical to do by hand.  Consequently, the HFGT toolbox has been validated by making sure that its results match the results found in a number of early HFGT publications\cite{Farid:2006:IEM-C02,Farid:2007:IEM-TP00,Khayal:2017:ISC-J35,Farid:2015:SPG-J17,Farid:2015:ETS-J25,Farid:2016:ETS-J27,vanderWardt:2017:ETS-J33,Abdulla:2015:EWN-C53} where either manual or semi-automated methods were used.   A wide variety of published test cases were tested so as to ensure that all of the HFGT toolbox data structures were instantiated and all of the HFGT toolbox methods were called. 

\section*{Petri Net Visualization}
The HFGT toolbox is then used in the Petri net visualization GUI.  This GUI takes the system modeled by the HFGT toolbox along with a scheduled event list and visualizes the flow of operands through the system via a Petri net.  To visualize the system in a Petri net additional GPS locations and initial token counts need to be added to the input XML.  To visualize the flow of operands through the system a scheduled event list needs to be created to describe the firing of DOFS and path of tokens.  

\section*{Creating a HFGT Petri net XML Input File}
The XML required by the Petri net GUI for visualizing the HFG takes the same form and holds the same attributes as the XML used solely to create a HFG.  However, it requires the addition of several attributes to {\tt machines}, {\tt independent buffers} and {\tt processes}.  {\tt Machines} and {\tt independent buffers} require the addition of initial token counts while all {\tt processes} require the addition of initial token counts, GPS offset, and the process duration time.  If the process is instantaneous of continuous in nature then the duration is set to 0.  These are described here in turn.  

\begin{figure}[!ht]
\centering
\includegraphics[width=\linewidth]{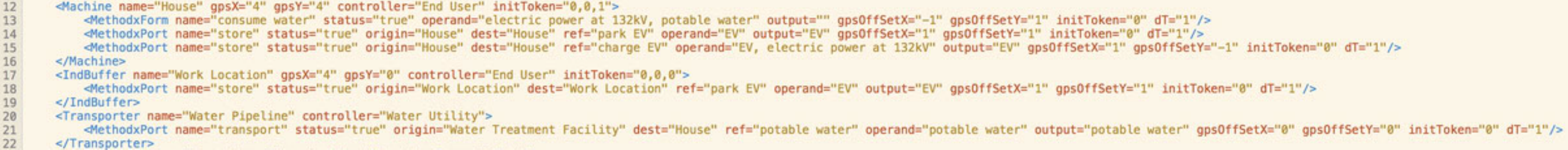}
\caption{An example HFGT Toolbox input XML input file that is PetriNet GUI compatible}
\label{fig:Trimetrica_Dummy_PN}
\vspace{-0.15in}
\end{figure}

\begin{trivlist}
\item \textbf{Machine}: This XML element is representative of a “Transformation Resource” (M) shown in Figure \ref{fig:LFES_ResourceArchitecture}. The attribute "initTokens" needs to be added to describe the number of tokens present at the place when the system is initialized.  The illustrative example in Figure \ref{fig:Trimetrica_Dummy_PN} shows the addition of "initTokens" to a transformation resource.
\item \textbf{IndBuffer}: This XML element is representative of an “Independent Buffer” (B) as shown in Figure \ref{fig:LFES_ResourceArchitecture}. The attribute "initTokens" needs to be added to describe the number of tokens present at the place when the system is initialized.  The illustrative example in Figure \ref{fig:Trimetrica_Dummy_PN} shows the addition of "initTokens" to a transformation resource.
\item \textbf{MethodxForm}: This XML element is used to describe the ability of a transformation resource to transform an operand. As shown in Figure \ref{fig:LFES_ResourceArchitecture} and as reflected in the illustrative example in Figure \ref{fig:Trimetrica_Dummy_PN}, this element is nested within the Machine (i.e. transformation resources) element. This element is also represented as a transition in the Petri net.  It thus requires the addition of attributes "gpsOffSetX" and "gpsOffSetY" to describe the locational GPS offset the transition will be displayed. It also requires the addition of the attribute "initTokens" to describe the number of tokens starting in the transition.  Additionally it requires the attribute "dT" to describe the duration of time the process takes to complete.  The illustrative example in Figure \ref{fig:Trimetrica_Dummy_PN} shows the addition of "gpsOffSetX", "gpsOffSetY", "initTokens", and "dT" to transformation processes.
\item \textbf{MethodxPort}:  As shown in Figure \ref{fig:LFES_ResourceArchitecture}, the operations involving holding or transportation of an operand are universal to all of the physical resources. This element is also represented as a transition in the Petri net.  It thus requires the addition of attributes "gpsOffSetX" and "gpsOffSetY" to describe the locational GPS offset the transition will be displayed. It also requires the addition of the attribute "initTokens" to describe the number of tokens starting in the transition.  Additionally it requires the attribute "dT" to describe the duration of time the process takes to complete.  The illustrative example in Figure \ref{fig:Trimetrica_Dummy_PN} shows the addition of "gpsOffSetX", "gpsOffSetY", "initTokens", and "dT" to transportation processes.
\end{trivlist}

\section*{Creating a HFGT Petri Net Scheduled Event List Input File}
\begin{figure}[!ht]
\centering
\includegraphics[width=.5\linewidth]{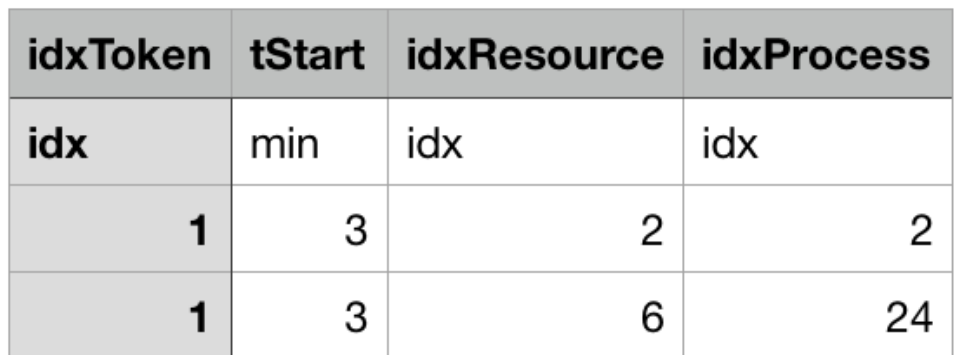}
\caption{An example Scheduled Event List input CSV file that is required but the PetriNet GUI.}
\label{fig:Trimetrica_Dummy_PN_SEL}
\vspace{-0.15in}
\end{figure}

The Scheduled Event List is required by the Petri net GUI to visualize the flow of (operand) tokens.  It uses a CSV file format comprised of several variables where each row represents an event.  An event is defined as the firing of a DOF on a specific token.  Each event thus requires the following variables to be defined with Figure \ref{fig:Trimetrica_Dummy_PN_SEL} providing an illustrative example of the scheduled event list.

\begin{trivlist}
\item \textbf{idxToken}: This variable in the ScheduledEventList specifies the specific token being operated on by the DOF.  This allows for the tracking of individual tokens to be tracked as they flow through the system.
\item \textbf{tStart}: This variable in the ScheduledEventList specifies the time at which the event begins.  
\item \textbf{idxResource}: This variable in the ScheduledEventList specifies the index of the resource at which the event fires.  The variable is later used to calculate the specific DOF being called by the event.
\item \textbf{idxProcess}: This variable in the ScheduledEventList specifies the index of the process that is used in the event.  The variable is later used to calculate the specific DOF being called by the event.
\end{trivlist}

\section*{Petri net Data Structures and Functions}

\begin{figure}[!ht]
\centering
\includegraphics[width=0.75\linewidth]{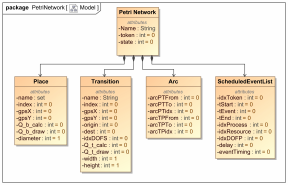}
\caption{Structure of myLFES on completion of XML2LFES}
\label{fig:myPetriNetwork}
\vspace{-0.15in}
\end{figure}

\begin{figure}[!ht]
\centering
\includegraphics[width=0.95\linewidth]{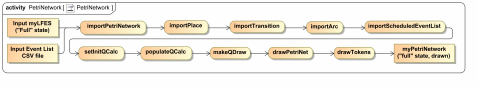}
\caption{A high-level activity diagram of the Petri Net Toolbox}
\label{fig:PN_Toolbox_overview}
\vspace{-0.15in}
\end{figure}

The \emph{Petri Net Visualization} uses MATLAB's object-oriented programming functionality to create generic Petri net class structure.   The primary purpose of the Petri net toolbox is to instantiate and subsequently populate the {\tt myPetriNetwork} structure from the data in the {\tt myLFES} structure and scheduled event list.  It's associated class diagram is shown in Figure \ref{fig:myPetriNetwork}.  The {\tt myPetriNetwork} stores all of the information from the scheduled event list and required information from the {\tt myLFES} structure. {\tt myPetriNetwork} then calculates all of the mathematical quantities for tracking the flow of operand tokens.
\noindent This {\tt myPetriNetwork} object is filled in by the constructor method of PetriNetwork when given a {\tt myLFES} object and a scheduled event List.  Once constructed several methods belonging to myPetriNetwork are called by the GUI to calculate the matrices describing the flow of tokens and used for visualizing the discrete events. Their sequence is shown in Figure \ref{fig:PN_Toolbox_overview}.  In brief, the {\tt import\_\_()} methods serve to import the {\tt myLFES} data and scheduled event list into their respective classes.  Then, the {\tt setInitQ()}, {\tt populateQCalc()}, {\tt makeQDraw()} methods calculate the matrices used to track the flow of tokens mathematically and visually.  Finally, the {\tt draw\_\_()} methods create the graphical objects that are drawn on the GUI axis to visualize the Petri net.  Each of these methods is called in turn by the GUI leaving {\tt myPetriNetwork} in a "full" and drawn state.  Each method is now discussed in detail.  

\vspace{0.2in}
\subsection*{myPetriNetwork = PetriNetwork(myLFES, scheduledEventList)}
As previously mentioned, the {\tt PetriNetwork()} constructor method serves to import the input myLFES and scheduled event list file and create the {\tt myPetriNetwork} ($N$) data structure.  This functionality is achieved through the sequence of functions shown in Figure \ref{fig:PN_Toolbox_overview}. 

\begin{trivlist}
\item \textbf{myPetriNetwork = importPetriNetwork(myLFES, scheduledEventList)} : This function is the PetriNetwork constructor and consequently instantiates {\tt myPetriNetwork} ($N$).  The high-level integer attributes of {\tt myPetriNetwork} are initialized to zero and the sub-classes are initialized through their constructor methods.  

\item \textbf{place = importPlace(myLFES)} : This function is the Place constructor and consequently instantiates \newline{\tt myPetriNetwork.place} ($S$).  This constructor assigns information to the sub-object {\tt myPetriNetwork.place} ($S$) by \say{unpacking} {\tt myLFES.machines} ($M$)and {\tt myLFES.indBuffers} ($B$).
 
\item \textbf{transition = importTransition(myLFES)} : This function is the Transition constructor and consequently instantiates \newline{\tt myPetriNetwork.transition} (${\cal E}$).  This constructor assigns information to the sub-object {\tt myPetriNetwork.transition} (${\cal E}$) by \say{unpacking} {\tt myLFES.machines$\{m\}$}{\tt .methodsxForm}, {\tt myLFES.machines$\{m\}$}{\tt .methodsxPort}, \newline{\tt myLFES.indBuffers$\{b\}$}{\tt .methodsxPort}, and {\tt myLFES.transporters$\{h\}$}{\tt .methodsxPort}. All processes are equated to transitions in the Petri net. 

\item \textbf{arc = importArc(myPetriNetwork)} : This function is the Arc constructor and consequently instantiates {\tt myPetriNetwork.arc} ($M$).  This constructor assigns information to the sub-object {\tt myPetriNetwork.arc} ($M$) by \say{unpacking} the place's locations and the transition's origins and destinations.  The associated arcs linking places and transitions as subsequently formed designating the direct of token flows.

\item \textbf{scheduledEventList = importScheduledEventList(eventList, myLFES)} : This function is the ScheduledEventList constructor and consequently instantiates {\tt myPetriNetwork.scheduledEventList}.  This constructor assigns information to the sub-object {\tt myPetriNetwork.scheduledEventList} by \say{unpacking} the input {\tt ScheduledEventList.CSV} and mapping events to their associated DOFs in {\tt myLFES}.

\item \textbf{myPetriNetwork = setInitQCalc(myPetriNetwork)} : This function expands the size of the $Q_{b_{calc}}$ and $Q_{t_{calc}}$ matrices in the sub-objects {\tt myPetriNetwork.place} ($S$) and {\tt myPetriNetwork.transition} (${\cal E}$) respectively such that they match the number of unique timings in the scheduled even list

\item \textbf{myPetriNetwork = populateQCalc(myPetriNetwork, myLFES)} : This function populates the $Q_{b_{calc}}$ ($Q_S$) and $Q_{t_{calc}}$ ($Q_E$) matrices in the sub-objects {\tt myPetriNetwork.place} ($S$)and {\tt myPetriNetwork.transition} (${\cal E}$) respectively.  Through tracking the activation of DOFs the flow operands through the system is tracked by the $Q_{b_{calc}}$ ($Q_S$) and $Q_{t_{calc}}$ ($Q_E$) matrices.

\item \textbf{myPetriNetwork = makeQDraw(myPetriNetwork)} : This function makes the $Q_{b_{draw}}$ and $Q_{t_{draw}}$ matrices in the sub-objects {\tt myPetriNetwork.place} and {\tt myPetriNetwork.transition} respectively.  These matrices are used to draw the calculated amount of tokens in each drawn place and transition.

\item \textbf{myPetriNetwork = drawPetriNet(myPetriNetwork)} : This function creates the geometric objects that are drawn on the GUI main axis to visualize the places and transitions of the Petri net.

\item \textbf{newHandles = drawTokens(myPetriNetwork,number, color, handles)} : This function draws the tokens at each state of the Petri net.  It uses information from the GUI stored in the handles structure and uses the {\tt color} and {\tt number} check boxes to determine if the number of tokens should be displayed via color map, numerically, or both.

\end{trivlist}

\section*{Petri net GUI Operation}

\begin{figure}[!ht]
\centering
\includegraphics[width=0.95\linewidth]{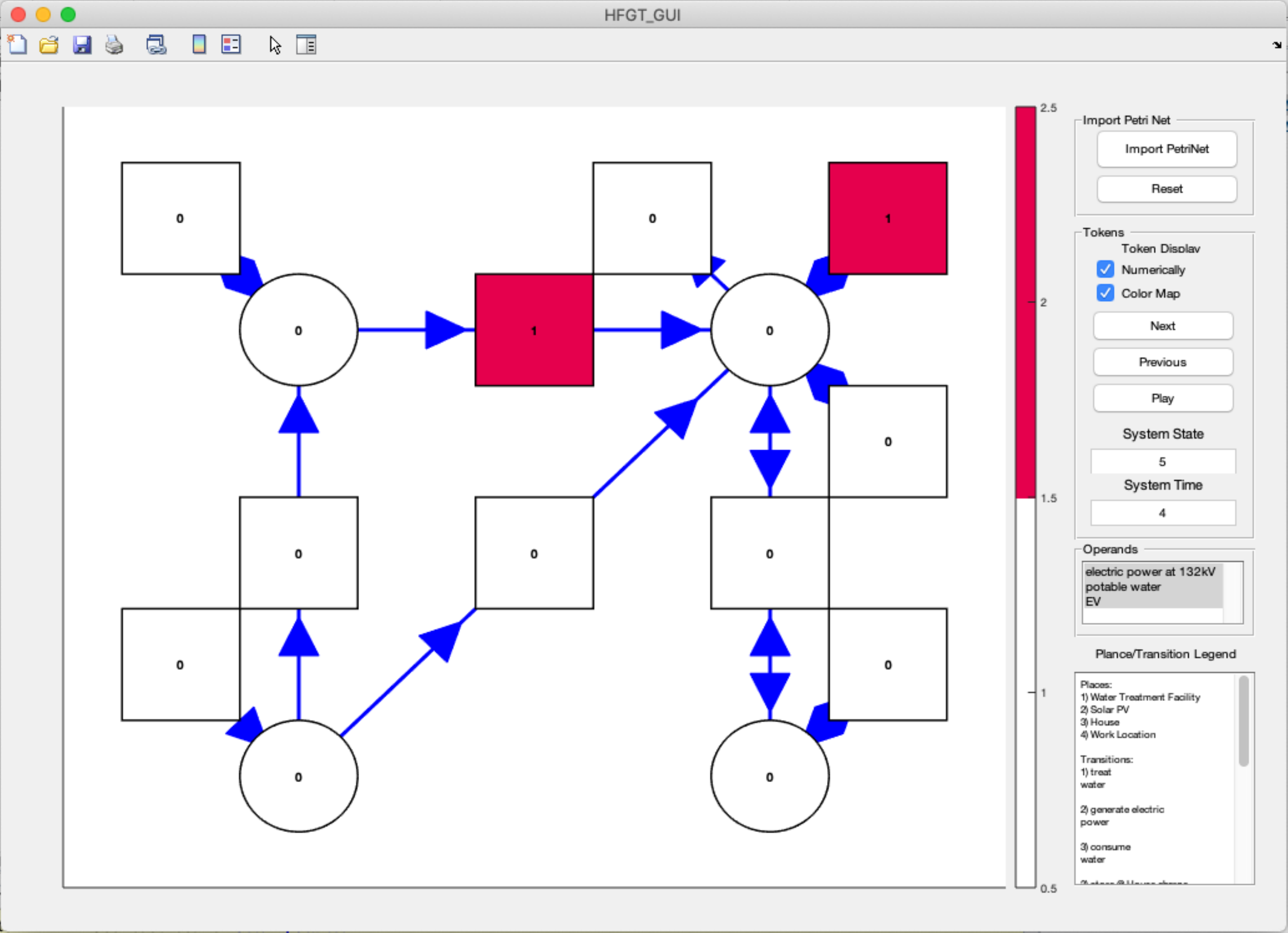}
\caption{An image of the Petri net GUI}
\label{fig:PN_GUI}
\vspace{-0.15in}
\end{figure}

The Petri net GUI uses the HFGT toolbox and Petri net structures to visualize the flow of operands through the HFGT system.  The GUI is shown in Figure \ref{fig:PN_GUI}.  The GUI components and operation are explained in turn.

\begin{trivlist}

\item \textbf{initAxes, Axis} : This is the main axis on which the Petri Net is displayed when plotted.  With the toolbar above it, the user is able to translate the plot and adjust the zoom of the plots view.  Additionally, when displaying tokens with color, a vertical color bar will appear on the right side to define the color scale.

\item \textbf{Import PetriNet, Button} : This button is a method for importing the Petri Net onto the main axis through a selection menu using dropdown menus and browse buttons.

\item \textbf{Reset, Button} : This button removes any Petri Net that has been imported into the GUI and clears the main axis of any graphic displayed on it.

\item \textbf{Numerically, Check Box} : This check box designates whether or not to display tokens numerically within each Place and Transition.

\item \textbf{Color Map, Check Box} : This check box designates whether or not to display tokens with a colormap\cite{Moreland:09:00} normalized over Places and Transitions.

\item \textbf{Next, Button} : This button advances the displayed Petri Net by incrementing the system state up one event.

\item \textbf{Previous, Button} : This button decrements the displayed Petri Net by decrementing the system state down one event.

\item \textbf{Play, Button} : This button continuously increments through the events of the displayed Petri Net.  Through a selection menu, the animation speed can be adjusted along with the animation being saved as an mp4 video file.

\item \textbf{System State, Edit Text} : This edit text box displays the index of the current state depicted by the main axis. Through entering a valid event index into the edit text box, the main axis will then display the desired state.

\item \textbf{System Time, Edit Text} : This edit text box displays the time associated with the current state depicted by the main axis. Through entering a time into the edit text box, the main axis will then display the state associated with a time closest to that is desired.

\item \textbf{Operands, Select Text} : This select text box displays the operands present in the system.  The main axis will only display tokens and transitions acting on the selected operands.  Any combination of operands from one to all can be selected.

\item \textbf{Place/Transition Legend, Inactive Edit Text} : This inactive edit text box displays the index and name of each place and transition depicted on the main axis as `$Index Name$'.

\end{trivlist}

\section*{Conclusion}
This paper serves as a user guide to a recently developed \emph{Hetero-functional Graph Theory Toolbox} and its associated \emph{Petri net Graphical User Interface}.  The toolbox and GUI are written in the MATLAB\cite{MATLAB:2019} language and has been tested with v9.6 (R2019a). It is openly available on GitHub together with a sample input XML file for straightforward re-use.  The paper details the syntax and semantics of the XML input, the myLFES (large flexible engineering system) data structure at the core of the toolbox and the functions used to construct and populate this data structure.  The paper also details the syntax and structure of the input scheduled event list, the myPetriNetwork, and the operation of the GUI.  The toolbox has been fully validated against several peer-review HFGT publications.  The development of a streamlined, computationally efficient, and openly-accessible toolbox and GUI that automates and visualizes the underlying mathematical operations of HFGT enables the broader scientific community to apply HFGT to a wide variety of highly interconnected and heterogeneous engineering systems.  Thus, this work enables several avenues for future research; particularly in the analysis, design, planning and operation of systems-of-systems.    
\vspace{-0.15in}

\section*{(2) Availability}
\vspace{0.5cm}
\section*{Operating system}

The toolbox and GUI are compatible with macOS 10.12 and higher, Windows 7, Windows 10, and all other operating systems that support MATLAB\cite{MATLAB:2019} v9.6 (R2019a).

\section*{Programming language}

The toolbox and GUI are written in the MATLAB\cite{MATLAB:2019} language and has been tested with v9.6 (R2019a).

\section*{Additional system requirements}

The toolbox requires 1.4MB on disc with 8GB of ram recommended.

\section*{Dependencies}

The Hetero-functional Graph Theory Toolbox is dependent on a XML\cite{Elliotte:2002:00} input file structure. The Hetero-Functional Graph Theory Toolbox also utilizes sparse tensors in its calculations and thus is dependent on the \href{www.tensortoolbox.org}{\textcolor{blue}{Tensor Toolbox for Matlab}}\cite{TTB_Sparse, TTB_Software}.

\section*{List of contributors}

Dakota J. Thompson is the lead author on this paper, secondary developer of the hetero-functional graph theory, and lead developer of the Petri net GUI.  Prabhat Hegde contributed to the writing this paper and contributed to the development of the hetero-functional graph theory toolbox.  Wester C.H. Schoonenberg is the lead developer of the hetero-functional graph theory toolbox and contributed to the writing of this paper.  Inas S. Khayal architected and managed the Petri net GUI and contributed to the development of hetero-functional graph theory.  Amro M. Farid architected the hetero-functional graph theory toolbox and Petri net GUI and managed the project including the writing of this paper.

\section*{Software location:}

{\bf Code repository} 
\begin{description}[noitemsep,topsep=0pt]
	\item[Name:] GitHub
	\item[Persistent identifier:] https://github.com/LIINES/HFGTToolbox
	\item[Licence:] tbc
	\item[Date published:] 28/08/20
\end{description}

\section*{Language}

The toolbox and GUI are written in the MATLAB\cite{MATLAB:2019} language and has been tested with v9.6 (R2019a).

\section*{(3) Reuse potential}

The development of a streamlined, computationally efficient, and openly-accessible toolbox and GUI that automates and visualizes the underlying mathematical operations of HFGT enables the broader scientific community to apply HFGT to a wide variety of highly interconnected and heterogeneous engineering systems.  Thus, this work enables several avenues for future research; particularly in the analysis, design, planning and operation of systems-of-systems. The GitHub repository will be updated appropriately as the software is modified or extended to support the work.

\section*{Competing interests}

The authors declare that they have no competing interests.

\bibliographystyle{unsrt}
\bibliography{LIINESHFGTToolbox.bib}

\end{document}